\documentclass[conference,twocolumn]{IEEEtran}
\usepackage{subfigure}
\usepackage{times,amsmath,epsfig}
\usepackage{latexsym,amssymb}
\usepackage{graphicx,amssymb,amstext,amsmath}
\usepackage{setspace}
\usepackage{amsmath}
\usepackage{amsthm}
\usepackage{eucal}
\usepackage{amssymb}
\usepackage{mathrsfs}
\usepackage{color}
\usepackage{url}
\usepackage{algorithm}
\usepackage{algpseudocode}
\usepackage{caption}
\usepackage{hyperref}
\newtheorem{theorem}{Theorem}
\newtheorem{definition}{Definition}

\def\qed{\hfill{$\Box$} \\}
\def\eqdef{\triangleq}

\def\12{\frac{1}{2}}

\def\lc{\left\lceil}   
\def\rc{\right\rceil}

\newfont{\bbb}{msbm10 scaled 500}

\newfont{\bb}{msbm10 scaled 1100}

 \IEEEoverridecommandlockouts
\begin{document}

\sloppy
\title{Delay Optimal Secrecy in Two-Relay Network} 

\author{
  \IEEEauthorblockN{Y.~Ozan Basciftci}
  \IEEEauthorblockA{Dep. of Electrical \& Computer Eng.\\
    The Ohio State University\\
    Columbus, Ohio, USA\\
    Email: basciftci.1@osu.edu} 
  \and
  \IEEEauthorblockN{C.~Emre Koksal}
  \IEEEauthorblockA{Dep. of Electrical \& Computer Eng.\\
    The Ohio State University\\
    Columbus, Ohio, USA\\
    Email: koksal@ece.osu.edu }
\thanks{This publication was made possible by NPRP grant  5 - 559 - 2 - 227 from the Qatar National Research Fund (a member of Qatar Foundation).} 
}


\maketitle
\begin{abstract}
We consider a two-relay network in which a source aims to communicate a confidential message to a destination while keeping the message secret from the relay nodes. In the first hop, the channels from the source to the relays are assumed to be block-fading and the channel states change arbitrarily -possibly non-stationary and non-ergodic- across blocks. When the relay feedback on the states of the source-to-relay channels is available on the source with no delay, we provide an encoding strategy to achieve the optimal delay. We next consider the case in which there is one-block delayed relay feedback on the states of the source-to-relay channels. We show that for a set of channel state sequences, the optimal delay with one-block delayed feedback differs from the optimal delay with no-delayed feedback at most one block.

\end{abstract}
\section{Introduction}
Delay required to communicate message $W$ from a source to a destination, is a key metric for communication networks. 
However, evaluating the optimal delay required to deliver the message in a network is not widely considered as it is very difficult to evaluate delay even in networks where no security constraint is imposed on a message. We consider a two-relay network with a secrecy constraint on a message, and do not make any assumption on the statistics of the source-to-relay channels, even on the existence of it.  We evaluate the minimum delay required to communicate the message to the destination reliably and securely, and find the algorithm that achieves it. 

\begin{figure}[t]
   \centering
   \includegraphics[width=0.30\textwidth]{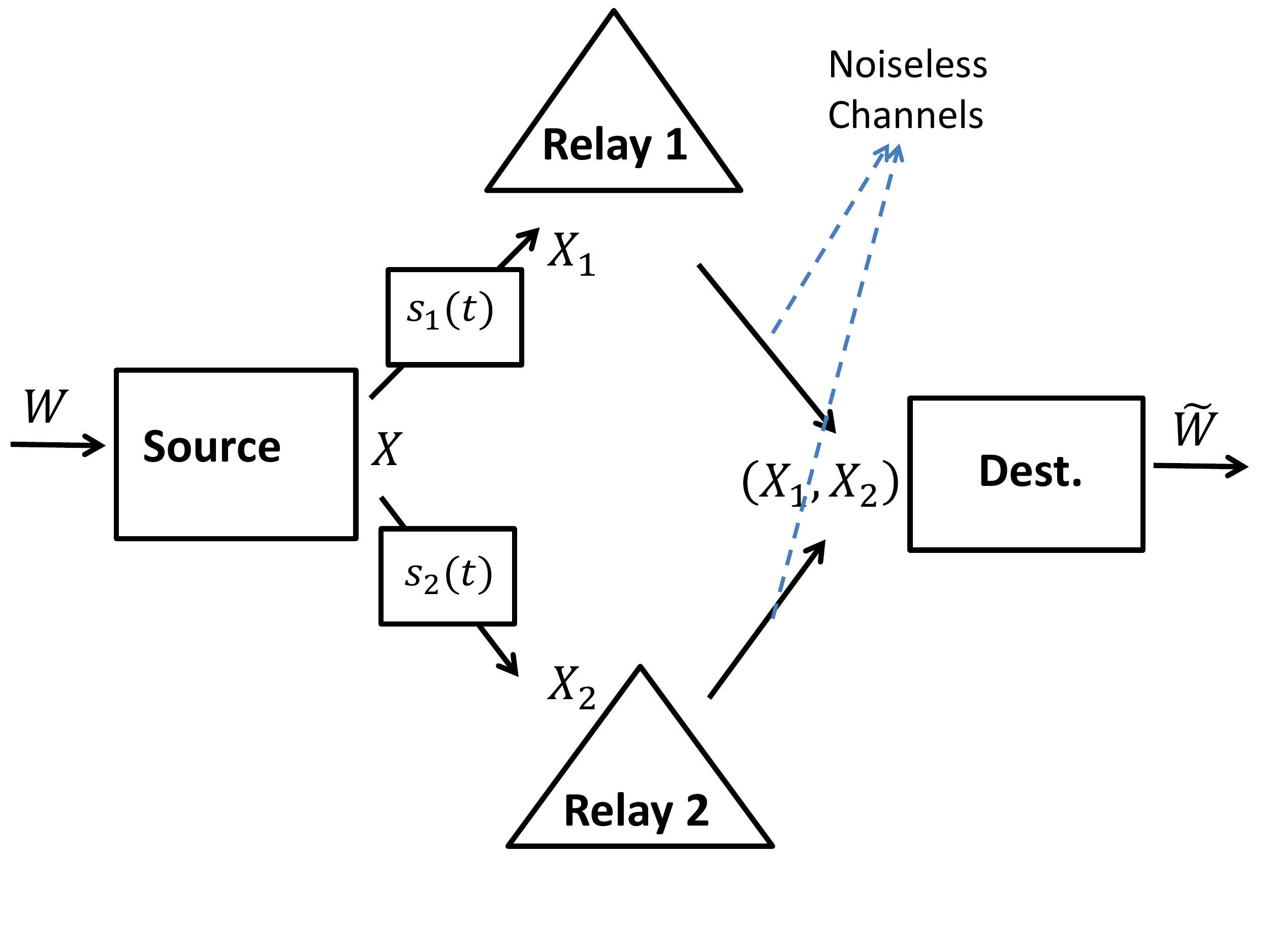}
   \caption{System Model}
   \label{dd}
 \end{figure}

The two-relay network we consider is depicted in Figure~\ref{dd}. The goal of the source is to communicate a \emph {finite} size message $W$ to the destination,
while keeping it secret from the relays. Source-to-relay 1 and source-to-relay 2 channels are assumed to be block erasure channels, and the states of relay channels
change one block to the next in an \emph{arbitrary manner}. Furthermore, we assume there is no direct 
channel from source to the destination, and both relay 1-to-destination and relay 2-to-destination channels are assumed to be noiseless.
We study this communication model under three set-ups each of which has a different channel state information (CSI) assumption: 1) Genie-aided CSI set-up: 
The source obtains the whole channel state sequence of the relay channels before the communication starts, 2) Zero-block-delayed CSI set-up: The source obtains the state of the relay channels at the beginning of a block, and 3) One-block delayed CSI set-up: The source obtains the state of the relay channel with a 1 block delayed feedback.
We evaluate the minimum number of channel blocks required to communicate message securely and reliably. 

The main challenge in our problem stems from the fact that since we delay with delay, we focus on the transmission of a message with  \emph{a finite and fixed size}. Hence, we cannot employ traditional asymptotic approaches \cite{wyner1975} to show the message is communicated
securely and reliably, since such approaches focus on large message sizes. To that end, we propose
encoding strategies for each CSI set-up to communicate the finite size message reliably and securely to the destination. Our contributions are as follows:
\begin{itemize}
\item We provide an encoding strategy to achieve the optimal delay of genie aided CSI set-up and optimal delay of zero-block delayed set-up $D_{\text{Zero-Block Delayed}}^*$. We observe that the optimal delays
of two set-ups are equal. 
\item We bound the optimal delay of the one-block delayed CSI set-up. 
 We show that the optimal delay of the one-block delayed CSI set-up differs from that of the zero-block delayed CSI set-up at most one block, if the source-to-relay 1 channel or the source-to-relay 2 channel does not experience an erasure on the channel block arriving after block $D_{\text{Zero-Block Delayed}}^*$. 
\end{itemize}

\emph{Related Work:}
In his seminal paper~\cite{wyner1975}, Wyner introduces the theoretical basis for information theoretic security for the point to point setting, where the adversary eavesdrops the communication between
the transmitter and the receiver. In~\cite{cai2002}, Cai and Yeung study the information theoretically secure communication of a message
in networks with general topologies, where the adversary can eavesdrop an unknown set of communication channels. The authors assume all the channels in the network have the same capacity.
In~\cite{ho2013}, the authors consider the same problem in~\cite{cai2002} in networks in which the channels do not need to have the same capacity. In~\cite{cai2002} and~\cite{ho2013}, the authors consider the communication channels as noiseless channels, whereas the source-to-relay 1 channel and the source-to-relay 2 channel are block erasure channels in our study.

In~\cite{soljanin2008}, the authors study information theoretically secure communication over \emph{noisy networks}, where each channel is assumed to be block erasure channel.
The authors provide upper and lower bounds to the secrecy capacity. In~\cite{diggavi2014}, the authors study  a secure communication over broadcast block erasure channel with channel state feedback at the end of each block. In both~\cite{soljanin2008} and \cite{diggavi2014}, the channel state changes from one block to the next in an independent and identically distributed fashion, whereas the channel state changes in an arbitrary manner in our study. Also, neither of \cite{soljanin2008} and \cite{diggavi2014} consider the delay of noisy networks, and both of them consider message size asymptotic regimes. The delay of a noisy network even without a secrecy constraint is very difficult to evaluate. We develop an encoding strategy for the genie aided CSI set-up and for the zero-block delayed CSI set-up, that achieves the minimum achievable delay of the two-relay network. For the one-block delayed CSI set-up, we provide a novel encoding strategy, and characterize the relation of the optimal delay of  the one-block delayed CSI set-up with that of the zero-block delayed CSI set-up. The encoding strategies we provide in the paper also keep the message secret from the relays without any assumption on the channel statistic.


\section{System Model and Problem Formulation}
We study the communication system illustrated in Figure~\ref{dd}. The source has a message $w\in\mathcal{W}$ to transmit to the destination over 2-relay network. The source-to-relay 1 and the source-to-relay 2 channel are block erasure channels. In the block erasure channel model, time is divided into discrete blocks each of which contains $N$ channel uses. The channel states are assumed to be constant within a block and vary from one block to the next in an arbitrary manner. Relay 1-to-destination and relay 2-to-destination channels are assumed to be error-free, i.e there is a wired connection between the relays and the destination. The observed signals at the relays and the destination in the $i$-th block are as follows:
\begin{align}
z_1^N(i)=&
\begin{cases}
x^N(i) &\mbox{ if } s_1(i) = 1\\
\emptyset& \mbox{ if } s_1(i) = 0
\end{cases}\\
z_2^N(i)=&
\begin{cases}
x^N(i) &\mbox{ if } s_2(i) = 1\\
\emptyset& \mbox{ if } s_2(i) = 0\\
\end{cases}\\
y^N(i)=&
\begin{cases}
x^N(i) &\mbox{ if } s_1(i) = 1 \text{ or } s_2(i) = 1\\
\emptyset& \mbox{ if } s_1(i) = 0 \text{ and } s_2(i) = 0
\end{cases}
\end{align}
where $x^N(i)\in\{0,1\}^N$ is the transmitted signal at $i$-th block, $z_1^N(i)$ is the received signal by the relay 1, $z_2^N(i)$ is the received signal by relay $2$, and $y^N(i)$ is the received signal by the destination at $i$-th block. With loss of generality, we assume that at each channel use, the source-to-relay $1$ channel and  the source-to-relay $2$ channel accept binary inputs, $\{0,1\}$. Channel states $s_1(i)$ and $s_2(i)$ denote the state of the source-to-relay $1$ channel and the state of the source-to-relay $2$ channel at $i$-th block, respectively. Equality $(s_1(i) = 1)$ denotes that  the source to relay $1$ channel is in on state, i.e there is no erasure at $i$-th block and $(s_2(i) = 0)$ denotes that the source to relay $1$ channel is in off state, i.e there is an erasure at $i$-th block.
Define $s(t)\eqdef [s_1(t),\;s_2(t)]$.

In this paper, we study the two-relay network in Figure~\ref{dd} under three set-ups each of which has a different channel state information (CSI) assumption. The set-ups are as follows: 1) Genie aided CSI set-up: The source knows whole state sequence, $\{s(t)\}_{t=1}^{\infty}$ before the communication starts,  2) Zero-block delayed CSI set-up:  The source  acquires the state of the channel block at the beginning of the corresponding block, 3) One-block delayed CSI set-up: The source  obtains the state of the channel block at the end of the corresponding block.

The source aims to send message $w\in\mathcal{W}=\{1,2,\ldots,2^{NR_s}\}$ to the receiver. By employing a $c(2^{NR_s}, DN)$, the encoder at the source maps message $w\in\mathcal{W}$ to a codeword $x^{DN}$, and the decoder at the destination, $d(\cdot)$ maps the received sequence $Y^{DN}$ to $\hat{w}\in\mathcal{W}$. The average error probability of a $c(2^{NR_s},ND)$ code is defined by
\begin{equation}
P_e^{DN} = 2^{-DNR_s}\sum_{w\in\mathcal{W}} \mathbb P(d(Y^{ND},\{s(t)\}_{t=1}^{\infty})\neq w| w\text{ was sent})
\end{equation}
The secrecy of transmitted message, $w$ is measured by the equivocation rates at relay 1 and relay 2, which are equal to the entropy rates of the transmitted message conditioned on  the observations of relay 1 and the observations of relay 2, respectively.
\begin{definition} \label{optimum_definition}Delay $D \eqdef D\left(R_s, \{s(t)\}_{t=1}^{\infty}\right)$ is said to be achievable if there exists a channel code $c(2^{NR_s}, ND)$ for which
\begin{align}
&P_e^{ND} = 0\nonumber\\
&\frac{1}{N}H\left(W|Z_1^{ND}, s^{D}\right) = R_s,\quad
\frac{1}{N}H\left(W|Z_2^{ND}, s^{D}\right) = R_s\nonumber
\end{align}
for any $N\geq 1$.
\end{definition}
The optimum delay, $D^{*}\left(R_s,\{s(t)\}_{t=1}^{\infty}\right)$ is defined to be the infimum of the achievable delays. Specifically,
$$D^{*} \eqdef D^{*}\left(R_s,\{s(t)\}_{t=1}^{\infty}\right) \eqdef \inf D\left(R_s, \{s(t)\}_{t=1}^{\infty}\right)$$

In this paper, our goal is to characterize optimum delay of genie-aided CSI, zero-block delayed CSI, and one-block delayed CSI set-ups. Delays $D_{\text{Genie-Aided}}^*$, $D_{\text{Zero-Block Delayed}}^*$, and $D_{\text{One-Block Delayed}}^*$ are referred to as the optimum delays of genie-aided CSI, zero-block delayed CSI, and one-block delayed CSI set-ups, respectively. Note that as stated in Definition~\ref{optimum_definition}, block length $N$ does not require to be infinite.  The delay results we give in Sections \ref{opt-delay} and \ref{delay3} are valid for any finite $N$.


\section{The Optimal Delay of Genie-Aided CSI and Zero-Block Delayed CSI Set-ups  }
\label{opt-delay}
In this section, we provide the optimal delay of the genie-aided CSI set-up and the optimal delay for zero-delayed CSI set-up. We show that the optimal delay of genie-aided CSI set-up is equal to the optimal delay of the zero-delayed CSI set-up.
\begin{theorem}\label{optimal}
The optimal delay of the genie-aided CSI set up is equal to the optimal delay of the zero-delayed CSI set-up. The optimal delay of the genie-aided CSI set up is  as follows:
\begin{align}
D_{\text{Genie-Aided}}^*  =D_{\text{Zero-Block Delayed}}^*=& \min d\\
\text{subject to }&I_\text{off-on}\left(d,s^d\right)\geq \lc R_s \rc\nonumber\\
&I_\text{on-off}\left(d,s^d\right)\geq \lc R_s \rc\nonumber \\
&d\in\mathbb Z_+ \backslash \{0\} \nonumber
\end{align}
where
\begin{align}
&I_\text{on-off}\left(d,s^d\right)\eqdef |\{t\in[1:d]: s_1(t)=1,\; s_2(t)=0\}|, \\
&I_\text{off-on}\left(d,s^d\right)\eqdef |\{t\in[1:d]: s_1(t)=0,\; s_2(t)=1\}| 
\end{align}
\end{theorem}\qed
Define an on-off block as a block on which the source-to-relay $1$ channel is in on state and the source-to-relay $2$ channel is in off state.  Define an off-on block, an on-on block, and an off-off block in a similar way. Theorem $1$ states that delay $D$ is achievable if and only if the source observes $\lc R_s\rc$ on-off blocks and $\lc R_s\rc$ off-on blocks until the end of block $D$, and the optimal delay is the minimum of the achievable delays.  The encoding strategy to achieve the optimal delay is provided in Algorithm~\ref{enc_alg1}. Note that Algorithm~\ref{enc_alg1} runs successfully for both the genie-aided CSI set up and the zero-delayed CSI set-up. Hence, the delay achieved with Algorithm~\ref{enc_alg1} is an upper bound to both set-ups. 
We next prove Theorem~\ref{optimal}, and start the proof by explaining Algorithm~\ref{enc_alg1} in detail.

\begin{proof}
We first prove that $D$ is achievable if $\lc R_s \rc \leq I_{\text{off-on}}(D,s^D)$ and 
$\lc R_s \rc \leq I_{\text{on-off}}(D,s^D)$. The achievability strategy depicted in Algorithm $1$ is as follows. Message $w$ is partitioned into $\lc R_s\rc$ sub-messages, $\{w_i\}_{i=1}^{\lc R_s\rc}$, i.e., $w = \left[w_1,\ldots, w_{\lc R_s\rc}\right]$, each of which except the last sub-message has $N$ bits. The last sub-message is padded with random bits so that it has $N$ bits.
For the secure transmission of message $w$, the source generates a set of keys $\{k_i\}_{i=1}^{\lc R_s\rc}$. For each $i\in \left[1:\lc R_s\rc\right]$, key $k_i\in\{0,1\}^N$ is picked from random variable $K_i$ that is uniformly distributed on $\{0,1\}^{N}$ and is independent from message $W$ and random variables $\{K_j\}_{j=1, j\neq i}^{\lc R_s \rc}$.
The source encrypts each sub message as $w^{'}_i =w_i \oplus k_i$. The source sends the encrypted sub-messages in on-off blocks, and sends the keys in off-on blocks. Specifically, at the beginning of block $t$, the source observes the channel state. If block $t$ is an on-off block, the source sends the next encypted sub-message, i.e., $x^N(t)=w_i \oplus k_i$. If  block $t$ is an off-on block, the source sends the next key, i.e., $x^N(t)= k_i$. In on-on blocks and off-off blocks, the source remains silent.

The secrecy analysis of Algorithm 1 is as follows. The equivocation analysis below stands for the secrecy analysis for relay 1.
\begin{align}
&H(W|Z_1^{ND},s^D)\label{eq1}\\
&= H\left(\{W_i\}_{i=1}^{\lc R_s \rc}|\{W_i\oplus K_i\}_{i=1}^{\lc R_s \rc},s^D\right)\\
&\stackrel{(a)}{=}\sum_{k=1}^{\lc R_s \rc}  H\left(W_k|\{W_i\oplus K_i\}_{i=1}^{\lc R_s \rc},\{W_i\}_{i=1}^{k-1},s^D\right)\\
&\stackrel{(b)}{=}NR_s \label{eq2},
\end{align}
where $(a)$ follows from the chain rule, and $(b)$ follows from the fact $W_k$ and $\left\{\{W_i\oplus K_i\}_{i=1}^{\lc R_s \rc},\{W_i\}_{i=1}^{k-1}\right\}$ are independent and from the fact $W_k$ is uniformly distributed on $\{0,1\}^N$. In a similar way with derivation (\ref{eq1}-\ref{eq2}), we can show that $H(W|Z_2^{ND},s^D) = NR_s$
\begin{algorithm}
\caption{Encoding strategy in Genie-Aided CSI and Zero-Block Delayed CSI set-ups}
\label{enc_alg1}
\begin{algorithmic}[1]
    \State $i \leftarrow 1$, $j \leftarrow 1$, $t \leftarrow 1$ 
    \While {$i\leq \lc R_s \rc$ or $j\leq \lc R_s \rc$}
      \If{$\left[s_1(t),\; s_2(t)\right] = [1,\;0]$ and $i\leq \lc R_s \rc$ }
        \State  $x^N(t)\leftarrow w_i \oplus k_i$
        \State $i \leftarrow i+1$
      \ElsIf{$\left[s_1(t),\; s_2(t)\right] = [0,\;1]$ and $j\leq \lc R_s \rc$ }
        \State  $x^N(t)\leftarrow k_j$
        \State $j \leftarrow j+1$
        \Else
        \State $x^N(t) \leftarrow \emptyset$
       \EndIf 
      \State t = t +1
	\EndWhile
	\State $D^{*}_{Zero-Block Delayed} \leftarrow t$, $D^{*}_{Genie Aided} \leftarrow t$
\end{algorithmic}
\label{alg1}
\end{algorithm}

We next prove that delay $D$ is achievable only if $\lc R_s \rc \leq I_{\text{off-on}}(D,s^D)$ and 
$\lc R_s \rc \leq I_{\text{on-off}}(D,s^D)$. Suppose that delay $D$ is achievable. From Definition~\ref{optimum_definition} and Fano's inequality, we have 
\begin{align}
&H(W|Y^{ND},s^{D}) = 0\label{fano}\\
&\frac{1}{N}H\left(W|Z_1^{ND}, s^{D}\right) = R_s \label{equ1}\\
&\frac{1}{N}H\left(W|Z_2^{ND}, s^{D}\right) = R_s
\end{align}

Then, we have the following derivation:
\begin{align}
&R_s = \frac{1}{N}H(W) \\
&\stackrel{(a)}{=} \frac{1}{N}H\left(W|Z_1^{DN},s^D\right)-\frac{1}{N}H\left(W|Y^{DN},s^D\right) \label{init1}\\
&         \leq \frac{1}{N}H\left(W|Z_1^{DN},s^D)-\frac{1}{N}H(W|Y^{DN},Z_1^{DN},s^D\right)\\
&         = \frac{1}{N}I\left(W;Y^{DN}|Z_1^{DN},s^D\right)\\
&         \stackrel{(b)}{\leq} \frac{1}{N}I\left(X^{DN};Y^{DN}|Z_1^{DN},s^D\right)\\
&       \stackrel{(c)}{\leq} \frac{1}{N}\sum_{t=1}^{D} H\left(Y^N(t)|Z_1^N(t),s(t)\right)\nonumber \\
&\quad\quad\qquad -H\left(Y^N(t)|Y^{(t-1)N},X^{DN},Z_1^{DN},s^D\right)\\
&         \stackrel{(d)}{=} \frac{1}{N}\sum_{t=1}^{D} H\left(Y^N(t)|Z_1^N(t),s(t)\right)\\
&        \stackrel{(e)}{=}  \frac{1}{N}\sum_{\{t: s_1(t)=0,\; s_2(t)=1\}}  H\left(Y^N(t)|Z_1^N(t),s(t)\right)\label{cond_zero}\\
&        \stackrel{(f)}{=}  \frac{1}{N}\sum_{\{i: s_1(t)=0,\; s_2(t)=1\}}  H\left(X^{N}(t)|s(t)\right)\label{cond1}\\
&     \stackrel{(g)}{\leq }I_{\text{off-on}}(D,s^D)\label{samp_space}
\end{align}
where $(a)$ follows from \eqref{fano} and \eqref{equ1}, $(b)$ follows from the fact that $W\rightarrow X^{DN}, Z_1^{DN} \rightarrow Y^{DN}$ forms Markov chain, $c$ follows from the fact that conditioning reduces the entropy, $(d)$ follow from the fact that $Y^N(t)$ is a function of $\left(X^{DN}, s^D\right)$. In~\eqref{cond_zero}, $(e)$ follows from the fact that $Y^N(t)=Z_1^N(t)$ if $\left[s_1(t),\; s_2(t)\right] = [1,\; 0]$, $\left[s_1(t),\; s_2(t)\right] = [1,\; 1]$, or $\left[s_1(t),\; s_2(t)\right] = [0,\; 0]$. Hence, $H\left(Y^N(t)|Z_1^N(t),s(t)\right) = 0$, if $\left[s_1(t), s_2(t)\right]\neq \left[0,\;1\right]$. In~\eqref{cond1}, $(e)$ follows from the fact hat $Y^N(t) = X^N(t)$ and $Z^N_1(t) = \emptyset$. In~\eqref{samp_space}, $(g)$ follow from the fact that $X^N(t)$ is a random variable whose sample space is $\left[1:2^{N}\right]$.

With a derivation similar to (\ref{init1})-(\ref{samp_space}), we find that $R_s \leq I_{\text{on-off}}(D,s^D)$. Hence, we conclude that if $D$ is an achievable delay, it has to satisfy constraints  $R_s \leq I_{\text{on-off}}(D,s^D)$ and $R_s \leq I_{\text{off-on}}(D,s^D)$. Note that these constraints imply that $\lc R_s \rc \leq I_{\text{on-off}}(D,s^D)$ and $\lc R_s\rc \leq I_{\text{off-on}}(D,s^D)$, since  $I_{\text{off-on}}(D,s^D)$  and $I_{\text{off-on}}(D,s^D)$ are integers.
\end{proof}
\section{On the Optimal Delay of One-Block Delayed Set-up}\label{delay3}
In this section, we provide lower and upper bounds for the optimal delay of the one block delayed CSI set-up. The tightness of the bounds depend on the number of the consecutive off-off blocks arriving after block $D_{\text{Zero-Block Delayed}}^*$. If the first block arriving after block $D_{\text{Zero-Block Delayed}}^*$ is on-on block, on-off block, or off-on block, the optimal delay of one-block
delayed CSI set-up differs from that of genie-aided CSI set-up at most one block.
\begin{theorem}\label{thm1block}
 The optimum delay of the one block delayed CSI set-up is bounded as follows:
\begin{equation}
D_{\text{Zero-Block Delayed}}^*\leq D_{\text{One-Block Delayed}}^* \leq D' \label{ineq2}
\end{equation}
where 
\begin{align}
D' \eqdef &\min d\label{delay_delayed}\\
\text{subject to } & D_{\text{Zero-Block Delayed}}^*< d\nonumber\\
               & s_1(d)=1 \text{ or } s_2(d)=1\nonumber\\
               & d\in \mathbb Z_+ \backslash \{0\}\nonumber
\end{align}
\end{theorem}\qed
Define an on block  as a block on which at least one of the source-to-channels is in the on state. Block $D'$ given in Theorem~\ref{thm1block} is the first on-block incoming after block $D_{\text{Zero-Block Delayed}}^*$. Algorithm~\ref{enc_alg2} provides an encoding strategy to achieve delay $D'$.  
We next provide the proof of Theorem~\ref{thm1block}

\begin{proof}
 We first explain Algorithm 2 and then prove the second inequality in \eqref{ineq2}. 
 Message $w$ is partitioned into $\lc R_s\rc$ sub-messages, $\{w_i\}_{i=1}^{\lc R_s\rc}$, i.e., $w = \left[w_1,\ldots, w_{\lc R_s\rc}\right]$.
In Algorithm 2, there are two phases which are key generation phase and data transmission phase.  At the beginning of block $t$, if either key queue at relay $1$ or key queue at relay $2$ are empty, the source enters into the key generation phase. The source transmits random bit sequence $r(t)\in\{0,1\}^N$ that is picked from random variable $R(t)\in\{0,1\}^N$ which is uniformly distributed on $\{0,1\}^N$ and independent from message $W$.  If block $t$ is an on-off block (resp. off-on block), transmitted random packet, $r(t)$ will not be heard from relay $2$ (resp. relay $1$) and will be stored at key queue at relay $1$ (resp. key queue at relay $2$)  as key $k_n^{(1)}$, i.e., $k_n^{(1)}=r(t)$ (resp.  as key $k_m^{(2)}$, i.e., $k_m^{(2)}=r(t)$). If block $t$ is in an on-on block, $r(t)$ will be heard by both relays. Hence, no keys will be generated at both relay $1$ and relay $2$.

At the beginning of block $t$, if both key queues at relay $1$ and $2$ are non-empty, the source enters into the data transmission phase. The source encodes next sub-message, $w_i$ as $x^N(t) =  w_i \oplus k^{(1)}_m\oplus k^{(2)}_n$, and transmits $x^N(t)$ in block $t$. If  block $t$ is on-off block, key $k_n^{(2)}$ is removed from the key queue at relay $2$, and key queue at relay $1$ remains same. Key $k_m^{(1)}$ is used to encode next sub-message $w_{i+1}$.

The source switches back and forth between the key generation and data transmission phases  as described above until all sub-messages are transmitted. Next, we prove the second inequality stated in \eqref{ineq2}. First define two variables $d_1(w_i)$ and $d_2(w_i)$. Variable $d_1(w_i)$ is the block on which sub-message $w_i$ is transmitted, when the source observes CSI at the beginning of each block and employs the encoding strategy in Algorithm $1$. Variable $d_2(w_i)$ is the block at the end of which the source is ready to send sub-message $w_i$, when the source observes CSI at the end of each block and employs the encoding strategy in Algorithm $2$, i.e., the key queue at relay 1 and key queue at relay 2 are non-empty at the end block $d_2(w_i)$. Specifically, the source sends sub-message $w_i$ on the first on-block incoming after block $d_2(w_i)$. Hence, the proof is complete if we show that $d_1(w_{\lc R_s\rc}) =d_2(w_{\lc R_s\rc})$

We  prove  statement $d_1(w_{\lc R_s\rc}) =d_2(w_{\lc R_s\rc})$ by induction.  First, we show that $d_1(w_1) =d_2(w_1)$. Since the source employing Algorithm \ref{alg1} transmits sub-message $w_1$ in block $d_1(w_1)$, block $d_1(w_1)$ is the first incoming block by the end of which the source observes at least one on-off block and at least one off-on block. Since the source starts the communication by sending random packets in Algorithm 2, key-queue at relay 1 and key queue at relay 2 will be non-empty at the end of block $d_1(w_1)$. Hence, at the end of block $d_1(w_1)$, the source employing Algorithm 2 is ready to send sub-message $w_1$ and $d_1(w_1)=d_2(w_2)$. Here, note that transmitted random packet in on-off block (resp., off-on block) will be  stored as a key in relay 1 (resp., relay 2)

Now assume that $d_1(w_{ i-1 }) = d_2(w_{ i-1})$ for any $1<i \leq \lc R_s\rc$. We next show that $d_1(w_i) = d_2(w_{i})$. For notational convenience define $I_\text{on-off}(i-1)\eqdef I_\text{on-off}\left(d_1(w_{ i-1 }), s^{d_1(w_{i-1})}\right)$  and $I_\text{off-on}(i-1)\eqdef I_\text{off-on}\left(d_1(w_{ i-1 }), s^{d_1(w_{i-1})}\right)$. Since the source employing Algorithm $1$ transmits sub-message $w_{i-1}$ in block $d_1(w_{i-1})$,  we have the following equality
\begin{equation}
i-1 = \min \left(I_\text{off-on}(i-1), I_\text{on-off}(i-1)\right)\label{min_operator}
\end{equation}
We first find the number of keys at key queue at relay $1$ and at key queue at relay $2$ at the end of block $d_1(w_{ i-1 })$. Assume w.l.o.g that by the end of block $d_2(w_{ i-1})$, the source employing Algorithm $2$ transmitted $v$ sub-messages at on-off blocks, $y$ sub-messages at off-on blocks, and $z$ sub-messages at on-on blocks, with $v+y+z = i-2$. The length of key queue at relay 1 at the end of block $d_2(w_{ i-1})$, $l_1(d_2(w_{ i-1}))$ is derived as follows:
\begin{align}
&l_1(d_2(w_{ i-1}))\stackrel{(a)}{=} I_\text{on-off}\left(i-1\right)-v-y-z\label{key_length0}\\
&\stackrel{(b)}{=} I_\text{on-off}\left(i-1\right)- \min \left(I_\text{off-on}(i-1), I_\text{on-off}(i-1)\right)+1\nonumber\\
&=\left[I_\text{on-off}\left(i-1\right)-I_\text{off-on}\left(i-1\right)\right]^++1 \label{key_length1}
\end{align}
where $(a)$ follows from the following facts: 1) In first $d_1(w_{i-1})$ blocks, the  source observes  $I_\text{on-off}\left(i-1\right)$ on-off blocks. In $\left(I_\text{on-off}\left(i-1\right)-v\right)$ of  $I_\text{on-off}\left(i-1\right)$ on-off blocks, the random packets are transmitted each of which are stored as a key at the key queue at relay $2$, 2) The keys  at key-queue at relay $1$, that are used in encoding sub-messages sent in $v$ on-off blocks are kept in the key queue, 3) The keys at key-queue at relay $1$, that are used in encoding sub-messages sent in $y$ off-on blocks and $z$ on-on blocks are removed from the key queue. In the derivation above, $(b)$ follows from the fact that $v+y+z = i-2$ and follows from Eq.~\eqref{min_operator}. With a similar derivation to Eq.~(\ref{key_length0})-(\ref{key_length1}), we can find the number of keys at key queue at relay $2$ at the end of block as $l_2(d_2(w_{ i-1}))=\left[I_\text{off-on}\left(i-1\right)-I_\text{on-off}\left(i-1\right)\right]^++1$

We prove  $d_1(w_i) = d_2(w_i)$ when $I_\text{on-off}\left(i-1\right) > I_\text{off-on}\left(i-1\right)$. The proof of   $d_1(w_i) = d_2(w_i)$ for case $I_\text{on-off}\left(i-1\right) \leq I_\text{off-on}\left(i-1\right)$ can be done similarly. Since  $I_\text{on-off}\left(i-1\right) > I_\text{off-on}\left(i-1\right)$, $l_1(d_2(w_{ i-1}))=I_\text{on-off}\left(i-1\right)-I_\text{off-on}\left(i-1\right)+1$, $l_2(d_2(w_{ i-1}))=1$, and block $d_1(w_i)$ is the first off-on block that arrives after block  $d_1(w_{i-1})$. 

The source employing Algorithm $2$ sends sub-message $w_{i-1}$ on the first on block arriving after block $d_1(w_{i-1})$. Let the block on which the source sends sub-message $w_{i-1}$ is off-on block. Then, the index of the off-on block is $d_1(w_{i})$. The length of the key queues at relay 1 and relay  2 at the end of block $d_1(w_{i})$ are $I_\text{on-off}\left(i-1\right)-I_\text{off-on}\left(i-1\right)$ and $1$, both of which are greater than zero. Hence, at the end of block $d_1(w_{i})$, the source employing Algorithm $2$ is ready to send message $w_i$, and $d_1(w_i) = d_2(w_i)$. Now let the block on which the source sends sub-message $w_{i-1}$ is either on-off block or on-on block and refer this block as block $s$. At the end of block $s$, the length of the key queue at relay 2 will be zero. Then, the source enters into key generation phase at the end of block $s$. The source keeps sending random packets until the end of the first off-on block arriving after block $s$. Note that the index of the first off-on block arriving after block $s$ is $d_1(w_{i})$. The random packet sent in block $d_1(w_{i})$ will be stored at key queue at relay 2 as a key and the lengths of key queue at relay $1$ and relay $2$ are non-zero at the end of block $d_1(w_{i})$.  Hence, at the end of block $d_1(w_{i})$, the source employing Algorithm $2$ is ready to send message $w_i$, and $d_1(w_i) = d_2(w_i)$.

\end{proof}
\begin{algorithm}
\caption{Encoding strategy in One-Block Delayed CSI set-up}
\label{enc_alg2}
\begin{algorithmic}[1]
\State $i\leftarrow 1$, $t\leftarrow 1$, $m\leftarrow 1$, $n\leftarrow 1$, $\text{Key-Queue1}\leftarrow 0$, $\text{Key-Queue2}\leftarrow 0$
\While {$i\leq \lc R_s \rc$} \Comment{$i$ is the sub-message index}
	\If{$\text{Key-Queue1}>0$ and $\text{Key-Queue2}>0$ }
      \State $\text{SendData} \leftarrow 1$
      \State $x^N(t)\leftarrow w_i \oplus k^{(1)}_m\oplus k^{(2)}_n$\Comment{Data transmission}
      \State $i\leftarrow i+1$
     \Else
      \State $\text{SendData} \leftarrow 0$ 
      \State $x^N(t)\leftarrow r(t)$ \Comment{Key generation}
     \EndIf

\Comment{Update the key queues at Relay 1 and 2 at the end each block}
\If{$\left[s_1(t),\; s_2(t)\right] = [1,\;0]$}
    \If{$\text{SendData} = 1$}
    \State $\text{Key-Queue2}\leftarrow \text{Key-Queue2}-1$
    \State $n\leftarrow n+1$
    \Else
     \State $\text{Key-Queue1}\leftarrow \text{Key-Queue1}+1$
    \EndIf
\ElsIf{$\left[s_1(t),\; s_2(t)\right] = [0,\;1]$}
    \If{$\text{SendData} = 1$}
    \State $\text{Key-Queue1}\leftarrow \text{Key-Queue1}-1$
    \State $m\leftarrow m+1$
    \Else
     \State $\text{Key-Queue2}\leftarrow \text{Key-Queue2}+1$
    \EndIf
\ElsIf{$\left[s_1(t),\; s_2(t)\right] = [1,\;1]$}
    \If{$\text{SendData} = 1$}
    \State $\text{Key-Queue1}\leftarrow \text{Key-Queue1}-1$
     \State $\text{Key-Queue2}\leftarrow \text{Key-Queue2}-1$
    \State $n\leftarrow n+1$
	\State $m\leftarrow m+1$
    \EndIf
\EndIf
\State $t\leftarrow t+1$ \Comment{$t$ is a block index}
\EndWhile
\State $ D^{*}_{One-Block Delayed}\leftarrow t$
\end{algorithmic}
\label{alg2}
\end{algorithm}
\section{Conclusion}
We study the minimum delay required to communicate the finite size message reliably to the destination in a two-relay network while keeping it secret from the relays, where source-to-relay channels are assumed to be block erasure channels. We provide an encoding strategy to achieve the optimal delay when the relay feedback on the states of the source-relay channels is available on the source with no delay, i.e., the source obtains the feedback at the beginning of a channel block.   Then, we consider the case in which there is an one-block delayed relay feedback on the states of the source-to-relay channels, i.e., the source obtains the feedback at the end of a block. We show that for a set of channel state sequences, the optimal delay with one-block delayed feedback differs from the optimal delay with no-delayed feedback at most one block.

\end{document}